          [2001/04/25 1.1 (PWD)]
\documentclass[a4paper,twocolumn]{esapub_new} % European paper
\usepackage{times}
\usepackage{natbib}
\usepackage{epsfig}

\title{\emph{RHESSI} observation of flare elements}
\author{Paolo C. Grigis}
\author{Arnold O. Benz}
\affil{Institute for Astronomy, ETH Z\"urich, 8092 Z\"urich, Switzerland}

\newcommand{\arcsec}{^{\prime\!\:\!\prime}}

\begin{document}

\maketitle

\begin{abstract}
RHESSI observations of \emph{elementary flare bursts} are presented.
These solar flare elements are distinct emission peaks of a duration
of some tens of seconds present in the hard X-ray light curves. They
are characterized by consistent soft-hard-soft spectral behavior,
which can be described in a quantitative way and compared which
predictions from acceleration models. A detailed analysis of hard 
X-ray images for an M5 class flare shows that elementary flare bursts
do not occur at distinct locations, but as twin X-ray sources move
smoothly along an arcade of magnetic loops. This observation apparently
contradicts the predictions of standard translation invariant
2.5-dimensional reconnection models.
\end{abstract}

\section{Introduction}

The hard X-ray emission from solar flares at energies higher than about
25 keV is strongly variable in intensity and often it is possible to
recognize several distinct emission peaks, lasting from a few seconds
to several minutes. Following \citet{deJager78}, we will call these
peaks in the light curve Elementary Flare Bursts (EFB) or, shortly,
flare elements. In the usual interpretation, non-thermal electrons
accelerated in the corona are responsible for the X-ray emission as
they precipitate into the lower, denser layers of the solar atmosphere
and emit bremsstrahlung at the footpoints (FP) of magnetic loops.
Therefore, the variation in the intensity of the X-ray light curves seen
during an EFB reflects a change in the flux of accelerated electrons,
which can be accounted for by a variation in the power of the accelerator,
or a change in the efficiency of the release of particles trapped in the corona.

The \emph{RHESSI} satellite \citep{lin02} provides light curves, spectra and
images. The interplay between these different sources of
information allows us to characterize the EFB in a much more detailed
way than purely temporal studies from light curves. The temporal evolution
of the spectra shows that the hardness of the photon spectrum changes during
a flare element, reaching its maximum at peak time. This behavior is called
Soft-Hard-Soft (SHS), and has been known for decades \citep{parks69}.
It has recently been studied in a more quantitative way by \citet{grigis04}
and \citet{grigis05}.

\begin{figure}
\epsfig{file=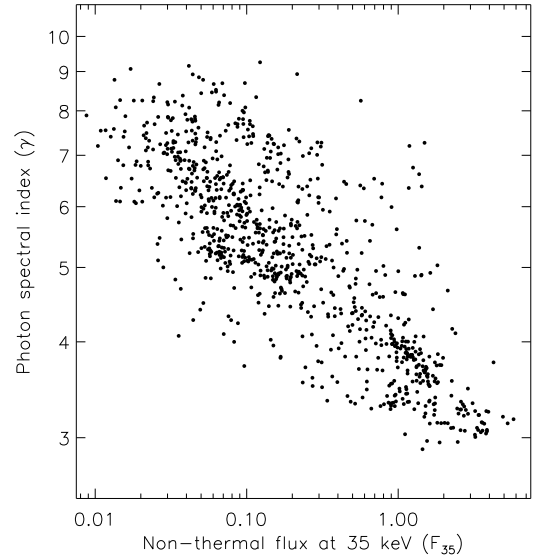,width=\linewidth}
\centering
\caption{Spectral index ($\gamma$) versus the fitted
non-thermal flux at 35 keV ($F_{35}$, given in photons
s$^{-1}$ cm$^{-2}$ keV$^{-1}$), for all 4 s intervals during 24 flares. \label{fig1}}
\end{figure}

Images of flare sources above 25 keV usually show one
or more sources, typically coming from footpoints of magnetic loops.
Two FPs on opposite sides of a magnetic neutral line are
expected in the standard model of eruptive flares (reviewed,
e.g., by Priest \& Forbes 2002). The rapid eruption of a cusp-shaped filament
enables the magnetic field to reconnect, driving particle acceleration
in field lines moving into the cusp. Electrons precipitate to the FPs of
these field lines in the chromosphere. In this scenario, one expects the
observed FP sources to drift apart as successive field lines are reconnected
at higher altitudes. This explanation fits the long-known outward motion
of H$\alpha$ ribbons parallel to the neutral line.
However, observations reveal a more complex picture, where source
motion can be as well parallel to the neutral line \citep{bogachev05}.
If the standard model of the solar flares holds for an EFB, it predicts 
systematic outward motions of the FP sources during a flare
element, while erratic jumps are expected between EFBs, as
different loop system are triggered into a phase of strong particle
acceleration and energy release.

Here we report observations of the general behavior of flare elements,
discussing them in more detail for one event showing continuous motion
along the ribbon of the arcade. The question of the specific motion
during flare elements is addressed in detail.

\section{Spectral evolution of the flare elements}

The spectral evolution for the EFBs is studied by producing high cadence
(4 s) spectral fittings to the incoming photons. The fitted model consist
of two components: an emission from an isothermal plasma and a
non-thermal power-law spectrum. The first dominates at low energies 
($<$ 25 keV). We are interested on the temporal development of the
second component, which is parameterized by its spectral index $\gamma$
and its normalization $F_{35}$ (that is, the strength of the photon
flux at 35 keV, given in photons s$^{-1}$ cm$^{-2}$ keV$^{-1}$).
Assuming a thick target model for the bremsstrahlung-producing
collisions of non-thermal electron with the cold ambient plasma,
the electron flux is still a power law with spectral index
$\delta= \gamma+1$ and its normalization can be easily computed.

\begin{figure}
\epsfig{file=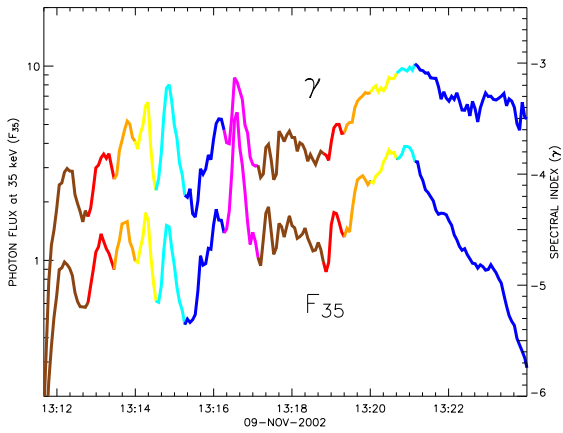,width=\linewidth}
\centering
\caption{Time evolution of the non-thermal photon flux at 35 keV ($F_{35}$)
and the spectral index ($\gamma$).\label{fig2}}
\end{figure}

The results of a set of fittings as the ones described above from a sample
of 24 uniformly selected flares in the time period February-November 2002,
ranging in size from GOES class M1 to X1, are shown in Fig. \ref{fig1}.
There is an evident trend correlating stronger flux with lower spectral index.
This shows an overall SHS effect. The effect however, holds not only as a general
trend between small and large flares \citep{battaglia05}, but also during single
flare elements.
This is exemplified by the flare of November 9, 2002. Figure \ref{fig2}
shows the evolution of the normalization of the non-thermal component
at 35 keV (lower curve) and the spectral index (upper curve). The colors have
been chosen to highlight the different flare elements. It is evident
that the SHS effect is more pronounced in each EFB than in the flare as a whole.

To be able to follow the evolution of the single EFBs in more details,
aiming for a more quantitative description of the effect, we switch
to the electron spectra. The quantitative analysis of 70 rise and decay
phases of flare elements belonging to the 24 selected flares
show that the electron spectra in a rise or decay
phase can be approximately described by having a common point of 
intersection, a \emph{pivot point}. The exact location of the pivot
point in the flux vs. energy diagram may be different for each
element, and also from the rise phase to the decay phase, but the
distribution of the pivot point energy $\epsilon_\mathrm{PIV}$
for all the rise and decay phases lies mainly between 10 keV and 30 keV
and is centered around 20 keV.
An asymmetry between the rise and decay phases is seen: the spectra
tend to harden with a faster rate in the rise phase than they
soften with decreasing flux.

\begin{figure}
\epsfig{file=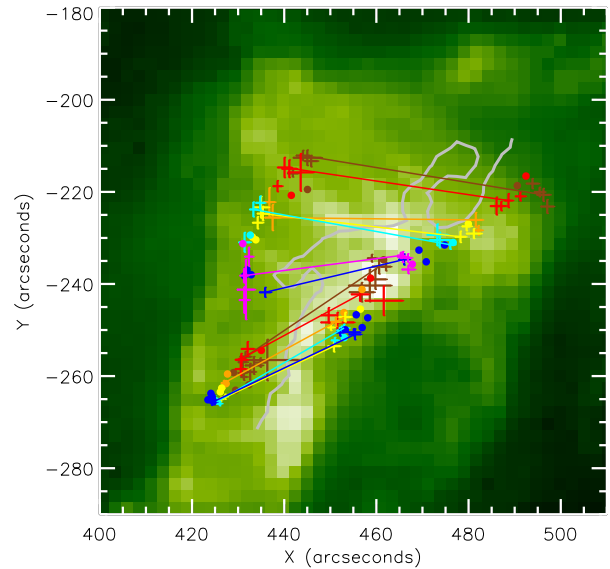,width=\linewidth}
\centering
\caption{%\emph{SOHO}/EIT 195 \AA\ image of postflare loops with the \emph{RHESSI}
%source positions superposed.
SOHO EIT 195 \AA\ image of postflare loops with the RHESSI 
HXR source positions superimposed. The positions of the 20--50 keV sources
from the CLEAN images are represented by crosses with arm lengths proportional
to the errors, positions from the PIXON images are given by circles.
Simultaneous FPs are connected and color coded according to the time intervals
defined in Fig. \ref{fig2}
\label{fig3}}
\end{figure}

The presence of a pivot point leads to an SHS effect at energies larger
than the pivot point energy and to an hard-soft-hard behavior at lower
energies. We note that the converse is not true: it is possible
to build a set of spectra showing the SHS effect at, say, 35 keV which
do not have a common point of intersection (for example, if all the
spectra were tangent to the same parabola).

A pivot point and SHS behavior can, at least approximately, be realized by
holding constant the number of electrons accelerated per second over a
threshold energy, or by the stochastic acceleration scenario proposed
by \citet{benz77} and \citet{brown85}. The authors are working on
a comparison of the spectral behavior with the prediction from more
sophisticated stochastic acceleration models.

\begin{figure}
\epsfig{file=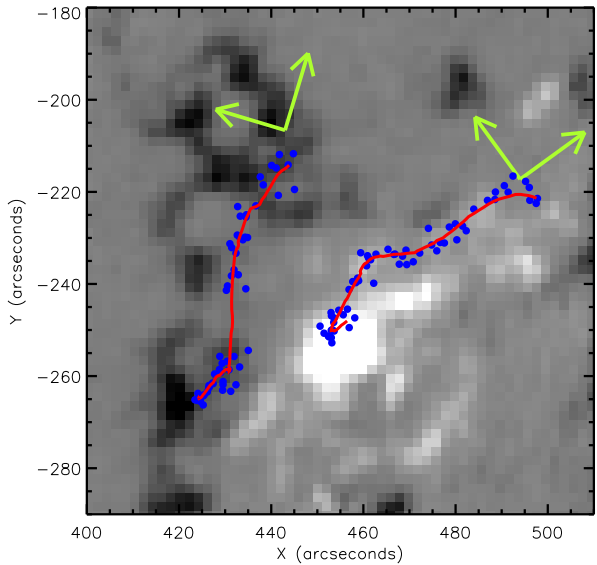,width=\linewidth}
\centering
\caption{\emph{SOHO} MDI magnetogram with the RHESSI PIXON source position
superimposed (blue circles). The red lines represent the smoothed trend line
for the source motions. The green arrows denote the \emph{parallel} and
\emph{perpendicular} directions with respect to the two ribbons. \label{fig4}}
\end{figure}

\section{Location of the flare elements}
The non-thermal electrons responsible for the power-law component
of the photon spectra are seen in hard X-ray images as loop footpoints.
We study in more detail the motion of the FP sources in the event
of November 9, 2002. This is an M5 class flare showing a pronounced post-flare
loop arcade in EIT images at 195 \AA. We produced a series of RHESSI images with
8 second cadence, using the CLEAN and the PIXON reconstruction algorithms.

\begin{figure}
\epsfig{file=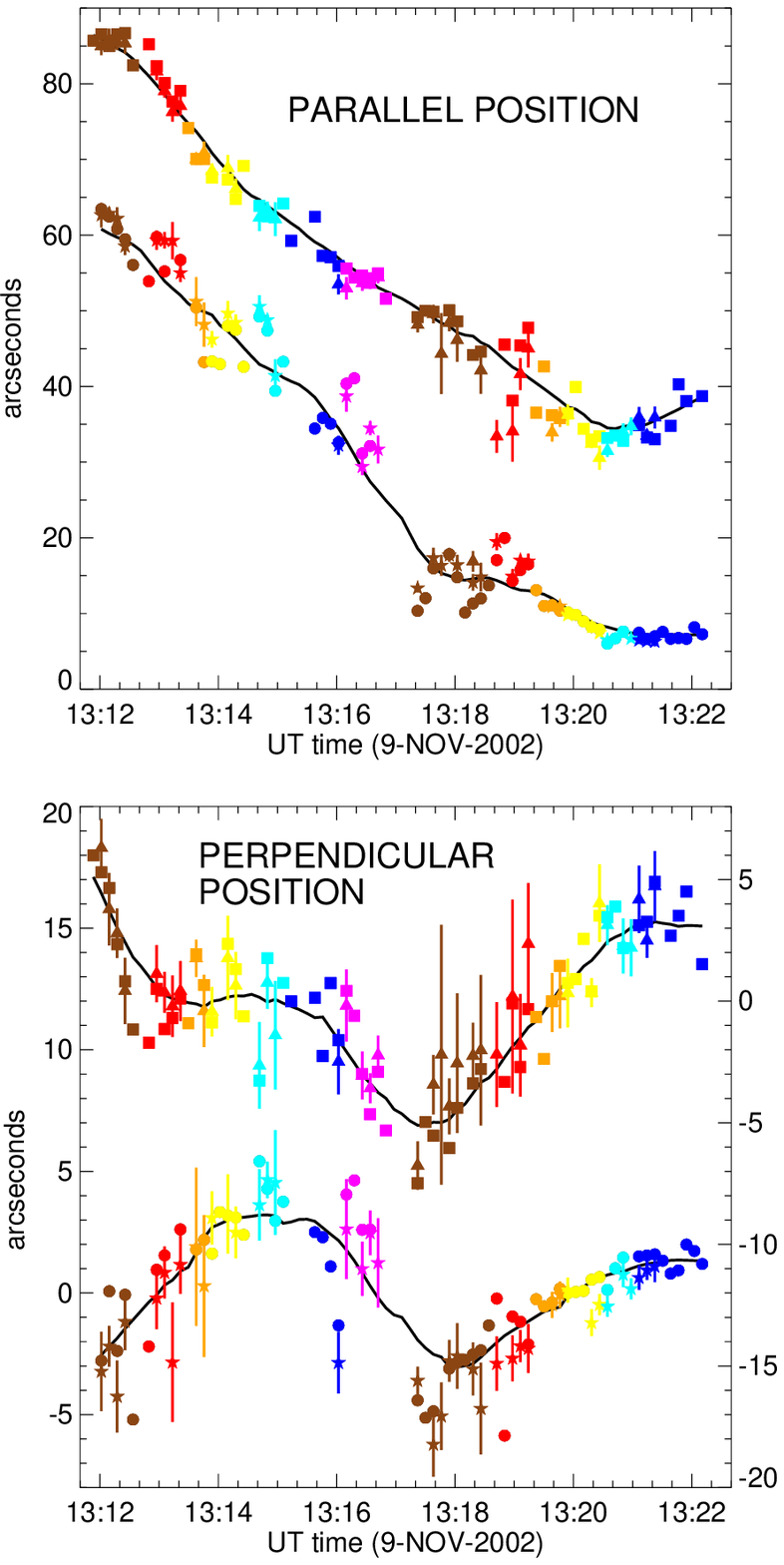,width=\linewidth}
\centering
\caption{Time evolution of the source positions relative to the trend lines.
The color code is the same as in Fig. 1, referring to the major subpeaks.
Triangles and stars with error bars refer to values derived using CLEAN,
squares and circles using PIXON, for the western and eastern FPs, respectively.
Top: The upper curve displays the parallel coordinates of the western FPs;
the lower curve shows the same, but for the eastern FPs. Bottom: Time evolution
of the coordinate perpendicular to the regression lines. The upper curve
refers to the western FP (scale on the right), the lower curve to the eastern
FP (scale on the left). Both panels show in black the averaged smoothed motion
for each FP (PIXON value), defining a new reference for detailed analysis
presented in Fig.~\ref{fighisto} \label{fig5}}
\end{figure}

The morphology of the hard X-ray images show two sources located at opposite
sites of the magnetic neutral line for most images. In some images only one
source is clearly defined. The source position was computed from a fitting of
a two-dimensional elliptical Gaussian to each visible source separately.

The source positions are shown on Fig.~\ref{fig3}, overlaid on an EIT image
at 195 \AA\ showing the loop arcade. They are also reported
on Fig.~\ref{fig4} as an overlay on an MDI magnetogram.
The red trend lines in Fig.~\ref{fig4} represent the position of the sources
smoothed with a moving average using a smoothing window corresponding to a
duration of 120 s, much longer than the individual duration of an EFB.
We now study the characteristics of flare element motion relative 
to the smooth trend line. Moreover,
we decompose the relative positions in two directions, \emph{parallel} and
\emph{perpendicular} to the arcade. These directions are obtained from coarse
fittings to the eastern and western source positions by straight lines. They
are also displayed in Fig.~\ref{fig4}.

\begin{figure}
\epsfig{file=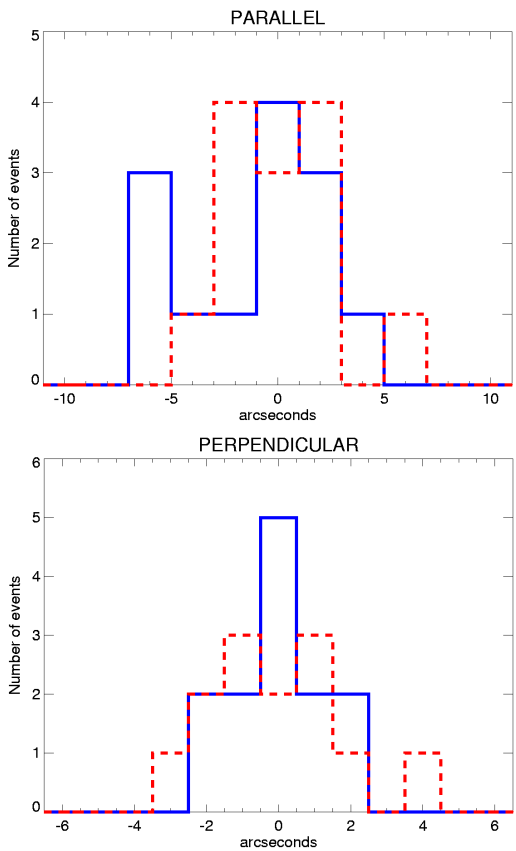,width=0.95 \linewidth}
\centering
\caption{Distribution of the average motions during a flare element
in perpendicular and parallel directions relative to the time
averaged trend curves. Eastern FPs are shown with continuous
blue lines, western FP with dashed red lines. \label{fighisto}}
\end{figure}

Do subpeaks show motions perpendicularly outward from
the ribbons, as expected from the standard reconnection model?
In Fig.~\ref{fig5} (bottom) this is not obvious, although the two FPs
are apparently moving relative to the regression line. Note,
however, that the lines are converging; thus, the effective FP
separation decreases. 
The standard reconnection model predicts outward FP motion
at a given place in the arcade. In order to look for such
systematic trends within HXR subpeaks, we took the parallel
and perpendicular components of the difference vector from
the smoothed source position to the observed PIXON positions.
For each subpeak, we averaged the positions occurring during
the first half and the second half of its duration. Then the difference
of the second minus the first half, $\Delta^\mathrm{POS}$, was calculated
for both eastern and western sources.

For EFBs produced by standard reconnection, one would expect
outward moving sources, thus $\Delta^\perp_\mathrm{POS}$
 being positive, at least
on average. Furthermore, the motion along the ribbons should
be stepwise and discontinuous with $\Delta^\parallel_\mathrm{POS}$
 being positive if each
EFB were a localized event. Fig.~\ref{fighisto} demonstrates that these
expectations are not satisfied during subpeaks of this flare. The
distribution of the average perpendicular motion during each
peak shown in Fig.~\ref{fighisto} has a mean $\Delta^\perp_\mathrm{POS}$
value of $0.0\arcsec \pm 0.4$
for the eastern FP and $0.2\arcsec \pm 0.5$ for the western FP
(the error is the standard error of the average). The mean value of the
relative parallel motion during the peaks is $-1.0\arcsec \pm 1.0$ for the
eastern FP and $0.4\arcsec \pm 0.7$ for the western FP.

The global motion along the arcade progresses with an average
velocity in the parallel direction of 63 km s$^{-1}$ for the
eastern FP and 55 km s$^{-1}$ for the western FP. The lower velocity
of the western FP is due to the fact that the last data points
have negative parallel velocities since they move backward
(Fig. 2). Averaging the absolute values of the parallel component
of the velocity, we get 65 km s$^{-1}$ for the western FP.
A speed of about 110 km s$^{-1}$ is maintained for 2 minutes in
the western FP at the beginning of the flare, while the data gap
and possible jump around 13:17 in the eastern FP position
requires 180 km s$^{-1}$.

\begin{figure*}
\epsfig{file=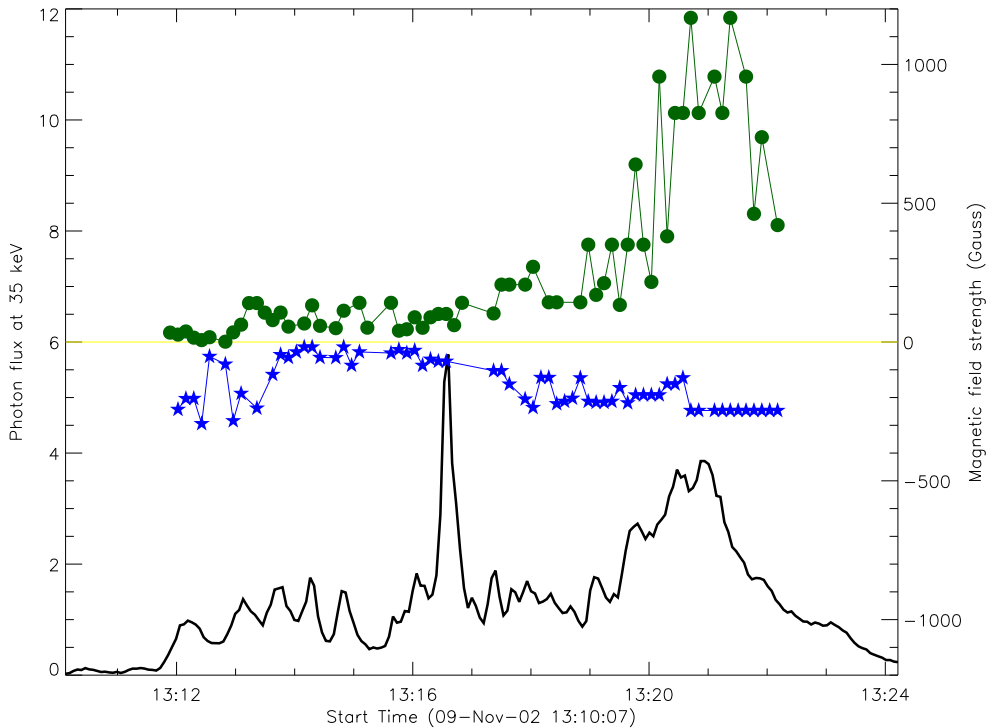,width=0.8\linewidth}
\centering
\caption{ Magnetic field strength encountered by the FP sources
(\emph{top}) and photon flux at 35 keV (\emph{bottom}) vs. time.
The photospheric field measured by SOHO/MDI (Fig. \ref{fig4}) shown
with green circles stands for the western FPs, the blue asterisks
represent the eastern FPs.\label{figmag}}
\end{figure*}

Since the ultimate energy source of a flare is provided by the
magnetic field, we expect that is possible to accelerate
more particles in regions of the arcade where stronger magnetic
fields are present. It is possible to measure photospheric
magnetic fields routinely, but we don't have easy access to
the values of coronal magnetic fields. Extrapolations are
possible, but neglects the effect of coronal currents.
Nevertheless, we can check whether there is a correlation
in our data between the photospheric magnetic field strength
and the hard X-ray flux (Fig. \ref{figmag}).

We take the field values from SOHO/MDI line-of-sight magnetograms
using the value of the MDI pixel nearest to the PIXON centroid
position of the footpoint sources.
For the western FP, there is a general trend of increase in the
magnetic field strength after 13:19, where the photon flux also
get larger. This is not seen for the eastern FP. This is
confirmed by the values of the cross correlation coefficients
$R_\mathrm{W}$ and $R_\mathrm{E}$ for the western and eastern
FP, respectively. We have
$R_\mathrm{W}=0.60{} ^{+0.08}_{-0.09} {}^{+0.14}_{-0.19}$
and $R_\mathrm{W}=0.19{}^{+0.12}_{-0.13} {}^{+0.23}_{-0.26}$,
where the errors indicate the interval for the 68.3\%
and 95.5\% confidence level, respectively.
However, if we compute the correlation coefficients separately
for each EFB and the build the average, we get lower values
${\overline{R}}_\mathrm{W}=0.24 \pm 0.17 \pm 0.36$ and
${\overline{R}}_\mathrm{E}=0.12 \pm 0.20 \pm 0.43$.
In Fig. \ref{figmag} we see that the strongest peak
happens in a region of weak magnetic field. The data
suggest that the effect of the magnetic field can be
seen in the overall, gradual progression of the flare,
but not for the impulsive flare elements. This indicates
that the plasma responsible for the local acceleration and
energy release in the loops may be dominated by region
of strong currents and/or turbulence and that the pre-flare
conditions of the magnetic field are far from a potential
state.

\section{Conclusion}

The flare elements seen very often in impulsive flares can be
identified in the lightcurves, but have also well defined spectral
behavior. The quantitative analysis of the behavior can be
compared with predictions from acceleration models, which
must be able to account not only for a general soft-hard-soft
behavior, but also for the presence of an approximate pivot
point in the electron spectra.
The consistent spectral behavior of the flare elements may suggest
that they constitute independent units representing separate
acceleration events during the flare.

The hard X-ray images for the event of November 9, 2002 confirm
this view partially as the different elements are found at
spatially distinct locations on the arcade. Different loops are
active at different times.
Surprisingly, the footpoints move smoothly along the two ribbons, however,
in contrast with the bursty evolution of the hard X-ray flux.
The parallel
source motions exclude the generally held notion of EFBs
being the modulation of a global reconnection process. Instead,
the temporal modulation of the HXR flux and spectral index
appear to be caused by a spatial displacement along the arcade.
This could be caused by some disturbance propagating smoothly
along the arcade, sequentially triggering a reconnection process
in successive loops of the arcade. The disturbance would have
to propagate with a speed in the range 50--150 km s$^{-1}$, much
lower than the Alfv\'en velocity.

In the impulsive phase of this flare, magnetic energy release
appears not in the form of a quasi-steady reconnection annihilating
antiparallel magnetic field and thus producing outward
moving FPs. The main flare energy release at a given position
in the arcade seems to last only a short time (order of a few
seconds) and moves along the arcade in a systematic manner.
The observed modulation of the HXR flux and the related
anticorrelation of the spectral index in each EFB appear to be
caused by spatial variations of the acceleration efficiency. The
temporal variations thus seem to be the result of a continuously
moving trigger propagating through variable conditions in the
arcade. The short lifetime of an FP at a given position shows
that particle trapping is not effective over timescales larger than
several tens of seconds.

The observed simple and systematic motions set this event
apart as a prototype for a type of HXR flare evolving along
the arcade. The FP motions of this flare clearly contradict the
expectations of the standard two-dimensional reconnection
model. The fact that we do not observe a systematic increase
(up to the instrumental limits) of the separation of the FPs does
challenge the idea that the reconnection points move upward,
and particles are accelerated in field lines successively farther
out during the main HXR emitting phase of the flare. A possible
interpretation is that the trigger releases the main energy stored
in a two-dimensional loop structure within seconds, without
noticeable FP motion, and moves on. Reconnection in the given
structure may still continue, but with HXR emission below
RHESSI sensitivity and at a much reduced energy release rate.

We thus propose a scenario in which a disturbance, probably
connected to the eruption of a filament, propagates along the
arcade like a burning fuse, sequentially triggering reconnection
and particle acceleration in the flare loops. The main HXR
emission from the FP reflects the propagation of this disturbance,
not the reconnection process at a given place in the
arcade. If the dominating emission is strong and short-lived,
the local conditions cause the observed temporal modulation.

The global evolution may be compatible with the standard
model of an eruptive flare, if one allows the filament to erupt
in such a way that one of its ends does not move while the
other starts to rise. In this scenario the reconnection process
spreads along the arcade until it reaches the end. The arcade
erupts in a manner similar to the opening of a zipper, where
the lower side run across the arcade and the upper side is the
filament. Future studies of HXR FPs in a large number of flares
may establish such a scenario and stimulate the development
of three-dimensional reconnection models needed to understand
these observations.

\section*{Acknowledgments}
The analysis of \emph{RHESSI} data at ETH Z\"urich is partially
supported by the Swiss National Science Foundation (grant
200020-105366). We thank the \emph{RHESSI} team for
their dedication and effort,  M. Battaglia for useful discussions,
and T. Wenzler for helping with the MDI data.

\end{document}